\documentclass[aps,prl,preprintnumbers,showpacs,nofootinbib]{revtex4}
\usepackage{graphicx}% Include figure files
\usepackage{dcolumn}% Align table columns on decimal point
\usepackage{bm}% bold math
\usepackage{subfig}
\usepackage{rotating} % Rotating tabl
\usepackage{multirow}
\usepackage{young}
\begin{document}
\newcommand{\newc}{\newcommand}
\newc{\ra}{\rightarrow}
\newc{\lra}{\leftrightarrow}
\newc{\beq}{\begin{equation}}
\newc{\eeq}{\end{equation}}
\newc{\barr}{\begin{eqnarray}}
\newc{\earr}{\end{eqnarray}}
%%%%%%%%%%%%%%%%%%%%%%%%%%%%%%%%%%%%%%%%%%%
\newcommand{\Od}{{\cal O}}
\newcommand{\lsim}   {\mathrel{\mathop{\kern 0pt \rlap
  {\raise.2ex\hbox{$<$}}}
  \lower.9ex\hbox{\kern-.190em $\sim$}}}
\newcommand{\gsim}   {\mathrel{\mathop{\kern 0pt \rlap
  {\raise.2ex\hbox{$>$}}}
  \lower.9ex\hbox{\kern-.190em $\sim$}}}
  \def\rpm{R_p \hspace{-0.8em}/\;\:}
%\preprint{APS/123-QED}
%\date{\today}
\title {Reduction of $SU_f(3)\supset SO(3)\supset A_4$-The scalar potential.}
%\author{ J. D. Vergados$^{1,2}$}
%\affiliation{1 Centre for the Subatomic Structure of Matter (CSSM), University of Adelaide, Adelaide SA 5005, Australia.}
%\affiliation{2 University of Ioannina, Ioannina, GR 45110, Greece.}
\author{ J. D. Vergados\footnote{Permanent address: University of Ioannina, Ioannina, Greece\\email:vergados@uoi.gr}}
\affiliation{ARC Centre of Excellence in Particle Physics at the Terascale and Centre for the Subatomic Structure of Matter (CSSM), University of Adelaide, Adelaide SA 5005, Australia}
\vspace{0.5cm}

\begin{abstract}
 We examine the possibility of the flavor symmetry being $SU_f(3)$  in which  the phenomenologically successful discreet symmetry $A_4$ is embedded. The  allowed quadratic and quartic potentials are constructed exploiting the full symmetry chain $SU(3)\supset SO(4)\supset A_4$. The spontaneous symmetry breaking in the  special case of the $SU_f(3)$  octet scalar fields is analyzed.
\end{abstract}
%\pacs{95.35.+d, 12.60.Jv}
%\pacs{ 13.15.+g, 14.60Lm, 14.60Bq, 23.40.-s, 95.55.Vj, 12.15.-y}
%%%%%%%%%%%%%%%%%%%%%%%%%%%%%%%%%%%%%%%%%%%%%%%%%%%%%%%%%%%%%%%%%%%%%
%\date{\today}
\maketitle
%\end{keyword}
%\end{frontmatter}
%%%%%%%%%%%%%%%%%%%%%%%%%%%%%%%%%%%%%%%%%%%%%%%%%%%%%%%%%%%%%%%%%%%%%
%%%%%%%%%%%%%%%%%%%%%%%%%%%%%%%%%%%%%%%%%%%%%%%%%%%%%%%%%%%%%%%%%%%%%
\section{Introduction}
For three generations of fermions the maximum horizontal symmetry is expected to be $SU(3)$ which contains as a subgroup the phenomenologically acceptable discreet symmetry $A_4$. It is known the the irreducible representations of $A_4$ are three singlets indicated as $\underline{1},\,\underline{1}^{'},\,\underline{1}^{''}$ and a triplet indicated as $\underline{3}$. The irreducible representations of $SU(3)$ are specified by two non negative integers $(\lambda,\mu)$. These are related to the young tableaux $\left [ f_1,f_2,f_3\right ],\,f_1 \geq f_2\geq f_3$ of $U(3)$ by $\lambda=f_1-f_2$ and $\mu=f_2-f_3$ In particle physics the $SU(3)$ representations are indicated merely by their dimension. This is not adequate for our purposes since there may be more than one irreducible representations with the same dimension.$(\mu,\lambda)$ is the adjoined (conjugate) of $(\lambda,\mu)$. Representations of the type $(\lambda,\lambda)$ are self-adjoined.

 The $SO(3)$ or $R(3)$ group is well known angular momentum theory. The $A_4$ is a discreet group isomorphic to the tetrahedral symmetry that has recently become popular in particle physics in connection with the neutrino mass. It is obviously a subgroup of $SO(3)$. Since, however, there exist three quark and lepton families,  one may suppose an $SU(3)$ flavor symmetry and would like to see the $A_4$ emerging as a subgroup of $SU(3)$. 
\section{The structure of  $A_4$}
 $S_4$ is the group permutations of 4 elements. It contains
\begin{enumerate}
\item The identity:
$$ C_1\LeftrightarrowΕ=\left ( \begin{array}{cccc}1\, 2\,3\,4\\1\, 2\,3\,4\end{array}\right )=(1\,1),\,(2\,2),\,(3\,3),\,(4\,4)$$
\item 6 transpositions
$$C_2\Leftrightarrow (12),\,(1\,3),\,(1\,4),\,(2\,3),\,(2\,4)$$
\item 8 cyclic permutations of length 3
$$C_3\Leftrightarrow(1\,2\,3),\,(3\,2\,1),\,(1\,2\,4),\,(4\,2\,1),\,(1\,3\,4),\,(4\,3\,1),\,(2\,3\,4),\,(4\,3\,2)$$
\item 6 cyclic permutations of length 4
$$C_4\Leftrightarrow(1\,2\,3\,4),\,(1\,2\,4\,3),\, (1\,3\,2\,4),\,(1\,3\,4\,2),\,(1\,4\,2\,3),\,(1\,4\,3\,2)$$
\item 3 products of two transpositions
$$C_5\Leftrightarrow(1\,2)(3\,4),\, (1\,3)(2\,4),\, (1\,4)(2\,3)$$
\end{enumerate}
We note that the 12 elements of the classes  $ C_1,C_3,C_5 $ form a group, $Α_4=C_1+C_3+C_5$ . The group  $Α_4$ is isomorphic to the tetrahedral group.\\
 with respect to $A_4$ the class $C_3$ splits into two classes:
$$C_{3α}\Leftrightarrow (1\,2\,3),\,(4\,2\,1),\,(4\,3\,2),\,(1\,3\,4),\quad C_{3β}\Leftrightarrow(1\,2\,4),\,(4\,2\,1),\,(4\,3\,1),\,(2\,3\,4)$$
So there exist 4 classes and, hence, 4 irreducible representations with dimensions constrained by:
\beq
\sum_{i=1}^3 \ell^2_i=12
\eeq
which has a unique solution:
\beq
\ell_1=1,\,\ell_2=1,\,\ell_3=1,\,\ell_4=3
\eeq
The three dimensional is the most important. We have :
\begin{itemize}
\item The identity element, $3\times3$.
\item The elements of class  $C_5$ commute and they can be simultaneously diagonalized. Since they are orthogonal with determinant equal to one they can be cast into the form:
$$ t_1=\left(
\begin{array}{rrr}
 1 & 0 & 0 \\
 0 & -1 & 0 \\
 0 & 0 & -1
\end{array}
 \right),\quad t_2=\left(
\begin{array}{rrr}
 -1 & 0 & 0 \\
 0 & 1 & 0 \\
 0 & 0 & -1
\end{array}
 \right),\quad t_3=\left(
\begin{array}{rrr}
 -1 & 0 & 0 \\
 0 & -1 & 0 \\
 0 & 0 & 1
\end{array}
 \right)
$$
This representation is often called real. The matrix  $t_2$ is sometimes denoted by$S^{'}$.   
\item the operator  $(1\,2\,3)$  induces the transformation $(x_1,x_2,x_3)\rightarrow (x_2,x_3,x_1)$ and  $(1\,3\,2)$  the transformation  $(x_1,x_2,x_3)\rightarrow (x_3,x_1,x_2)$. They are thus represented as:
$$
 (132)\Leftrightarrow a(0)=\left(
\begin{array}{lll}
 0 & 0 & 1 \\
 1 & 0 & 0 \\
 0 & 1 & 0
\end{array}
 \right),\quad 
(123)\Leftrightarrow b(0)=\left(
\begin{array}{lll}
 0 & 1 & 0 \\
 0 & 0 & 1 \\
 1 & 0 & 0
\end{array}
 \right).
$$
In terms of them we get
$$ a(i)=t(i) a(0) t(i),\quad b(i)=t(i) b(0) t(i),\quad i=1,2,3.
$$
In other words:
$$ 
α(1)=\left(
\begin{array}{rrr}
 0 & 0 & -1 \\
 -1 & 0 & 0 \\
 0 & 1 & 0
\end{array}
\right),\,α(2)=\left(
\begin{array}{rrr}
 0 & 0 & 1 \\
 -1 & 0 & 0 \\
 0 & -1 & 0
\end{array}
\right),\,α(3)=\left(
\begin{array}{rrr}
 0 & 0 & -1 \\
 1 & 0 & 0 \\
 0 & -1 & 0
\end{array}
\right)
$$
$$β(1)=\left(
\begin{array}{rrr}
 0 & -1 & 0 \\
 0 & 0 & 1 \\
 -1 & 0 & 0
\end{array}
\right),\,β(2)=\left(
\begin{array}{rrr}
 0 & -1 & 0 \\
 0 & 0 & -1 \\
 1 & 0 & 0
\end{array}
\right),\,β(3)=\left(
\begin{array}{rrr}
 0 & 1 & 0 \\
 0 & 0 & -1 \\
 -1 & 0 & 0
\end{array}
\right).
$$
The matrix  $b(2)$ is sometimes denoted by   $T^{'}$.
The elements  $T=(123)$  and $S=(14)(23)$ are the generators of the group. 

 Thus one can select a basis in which $T$  is diagonal. Indeed
$$T=U^+b(0)U=\left(\begin{array}{ccc}1&0&0\\0&\omega&0\\0&0&\omega^2 \\ \end{array}\right ),\,A=\sqrt{3} U=\left(\begin{array}{ccc}1&1&1\\ \omega&1&\omega^2\\ \omega^2&1&\omega\\ \end{array}\right ),\, \omega=e^{\frac{2 \pi i}{3}}$$
Then
$$U^+t_2U \rightarrow S=\frac{1}{3}\left(
\begin{array}{ccc}
 -1 & 2 & 2 \\
 2 & -1 & 2 \\
 2 & 2 & -1 \\
\end{array}
\right)
$$
$$U^+t_3U \rightarrow S_3=-1+\frac{2}{3}\left(
\begin{array}{ccc}
 1 & \omega  & \omega ^2 \\
 \omega ^2 & 1 & \omega  \\
 \omega  & \omega ^2 & 1\\
\end{array}
\right)
$$
$$U^+t_1U \rightarrow S_1=-1+\frac{2}{3}\left(
\begin{array}{ccc}
 1 & \omega^2  & \omega  \\
 \omega & 1 & \omega^2  \\
 \omega^2  & \omega  & 1\\
\end{array}
\right)
$$
\end{itemize}
\section{Application in $SU(3)\supset SO(3) \supset A_4$ chain}
We start with $SU(3)\supset SO(3)$ basis. The states in this chain are  specified by the $SU(3)$ quantum numbers $(\lambda,\mu)$  as well as the angular momentum quantum numbers $(L,M)$. The construction of these states is well known \cite{JDV68}. In general an additional quantum number $\kappa$ is needed,
$$|(\lambda,\mu)\kappa,L,M.$$
It is not easy to construct an operator, which is made up from the generators of $SU(3)$ and commutes with $L^2$ and $L_z$, with eigenvalues that specify $\kappa$.    So, whenever this additional quantum number is needed to remove   $L-$ degeneracy, we will  use an algorithm already developed \cite{JDV68}.

 The reduction of the $SO(3)$  representations to $A_4$  has been considered in the past,  with elegant but  abstract and rather obscure  for most physicists mathematical techniques  \cite{BerGros09,Ovrut78,Elesi96}. We will try to do this in a more direct  way motivated by physics. This way we will get a handle on how to construct the $SU(3)\supset SO(3)\supset A_4$ invariants and break the $SU(3)$ down to $A_4$

From the character tables of $A_4$ and $SO(3)$ one can show that the reduction of $SO(3)\supset A_4$. Since, however, $A_4$ is isomorphic to the tetrahedral symmetry $T$ it is adequate to consider th the well known expression \cite{JDV91} involving the reduction for $SO(3)\supset T$:
\beq
D^0=\Gamma^1,\,D^1=\Gamma^4,\,D^2=\Gamma^2\oplus\Gamma^3\oplus\Gamma^4,\,D^3=\Gamma^1\oplus2\Gamma^4.
\eeq
So the $L=1$ states are purely triplets, the $L=3$ states are linear combination of the $A_4$ invariant $\underline{1}$ and two triplets, while the $L=2$  states are linear combination of the two singlet non-invariant states  $\underline{1}'$ and s $\underline{1}''$ as well as one triplet.

It is thus obvious the $L=1$ transforms as $A_4$ triplet. The meaning of the tensors will become clear below when we consider the angular momentum states embedded in $SU(3)$. The results are  as follows:
\beq
|L=1,M\rangle \Leftrightarrow t^S_{M}=\left [ \left [{\bf p}\otimes{\bf q}\right  ]^1\otimes{\bf r}\right ]^1_M,\,M=\pm 1,0\mbox{ (symmetric triplet)}
\eeq
or
\beq
|L=1,M\rangle \Leftrightarrow t_{M}=\left [ \left [{\bf p}\otimes{\bf q}\right  ]^0\otimes{\bf r}\right ]^1_M,\,M=\pm 1,0.
\eeq
\beq
|L=2,M=\pm 1,0 \Leftrightarrow t^A_{M}=\left[\left [{\bf p}\otimes{\bf q}\right ]^1\otimes{\bf r}\right ]^2_M,\,M=\pm 1,0\mbox{ (antisymmetric triplet)}
\eeq
\beq
|L=2,M=2\rangle=\frac{1}{\sqrt{2}}\left(\underline{1}'+\underline{1}''\right),\,|L=2,M=-2\rangle=\frac{1}{\sqrt{2}}\left(\underline{1}'-\underline{1}'' \right),
\eeq
where $\underline{1}'$ and  $\underline{1}''$ are the two $A_4$ singlets, which are non invariant under $A_4$.
\beq
|L=3,M\rangle \Leftrightarrow t^S_{M}=\left [ \left [{\bf p}\otimes{\bf q}\right  ]^2\otimes{\bf r}\right ]^3_M,\,M=\pm 1,0\mbox{ (symmetric triplet)}
\eeq
The components of the other triplet are:
\beq
t^A_{1}=\frac{1}{\sqrt{2}}\left (\left [ \left [{\bf p}\otimes{\bf q}\right  ]^2\otimes{\bf r}\right ]^3_2+\left [ \left [{\bf p}\otimes{\bf q}\right  ]^2\otimes{\bf r}\right ]^3_{-2}\right )
\eeq
\beq
t^A_{-1}=\frac{1}{\sqrt{2}}\left (\left [ \left [{\bf p}\otimes{\bf q}\right  ]^2\otimes{\bf r}\right ]^3_2-\left [ \left [{\bf p}\otimes{\bf q}\right  ]^2\otimes{\bf r}\right ]^3_{-2}\right )
\eeq
\beq
t^A_{0}=\frac{1}{\sqrt{2}}\left (\left [ \left [{\bf p}\otimes{\bf q}\right  ]^2\otimes{\bf r}\right ]^3_3-\left [ \left [{\bf p}\otimes{\bf q}\right  ]^2\otimes{\bf r}\right ]^3_{-3}\right )
\eeq
where the antisymmetry is now with respect to conjugation.
Finally the invariant singlet is now given as 
\beq
\underline{1}=\frac{1}{\sqrt{2}}\left (\left [ \left [{\bf p}\otimes{\bf q}\right  ]^2\otimes{\bf r}\right ]^3_3+\left [ \left [{\bf p}\otimes{\bf q}\right  ]^2\otimes{\bf r}\right ]^3_{-3}\right )
\eeq
Thus we get
\barr
|l=3,M=3\rangle&=&\frac{1}{\sqrt{2}}\left ( \underline{1}+t^A_0\right),\,
|L=3,M=-3\rangle=\frac{1}{\sqrt{2}}\left ( \underline{1}-t^A_0\right),\,
|L=3,M=2\rangle=\frac{1}{\sqrt{2}}\left (t^A_1 +t^A_{-1}\right),\nonumber\\
|L=3,M=-2\rangle&=&\frac{1}{\sqrt{2}}\left (t^A_1 -t^A_{-1}\right),\mbox{ and }
|L=3,M\rangle=t^S_M,\,M={0,\pm 1}
\earr

 \subsection{Expansion of the regular and decuplet  $SU(3)$ representations}
 If each particle is in the fundamental $SU(3)$ representation $1,m$ the possible three particle states are $[2\times1]^3$, $[2\times1]^2$, $[2\times1]^1$, $[0\times1]^1$, $[1\times1]^2$, $[1\times1]^1$ and $[1\times1]^0$, where the indicated numbers  designate the angular momentum of the two particle state, the angular momentum of the third particle and, as an upper index, the total angular momentum of the thee particles respectively . The two particle states are designated in a similar manner $[1\times1]^L\,L=0,1,2$. 

The $SU(3)$ states are: i) $(\lambda,\mu)=(3,0)$, which corresponds to  the completely symmetric or $\underline{10}$-dimensional representation, ii) the regular $(\lambda,\mu)=(1,1)$,  $\underline{8}$ or octet and iii) the  $(\lambda,\mu=(0,0)$, singlet or completely antisymmetric representation. It is now straightforward  to expand the $SU(3)$ $\lambda,\mu  $ states in terms of the above $A_4$ basis. 
The results are as follows:
\begin{enumerate}
\item the $\lambda,\mu=(3,0)$ or decuplet $\underline{10}$ representation.
\barr
|(3,0)l=3,M=3\rangle&=&\frac{1}{\sqrt{2}}\left ( \underline{1}+t^A_0\right),\,
|(3,0)L=3,M=-3\rangle=\frac{1}{\sqrt{2}}\left ( \underline{1}-t^A_0\right),\nonumber\\
|(3,0)L=3,M=2\rangle&=&\frac{1}{\sqrt{2}}\left (t^A_1 +t^A_{-1}\right),
|(3,0)L=3,M=-2\rangle=\frac{1}{\sqrt{2}}\left (t^A_1 -t^A_{-1}\right),\nonumber\\
|(3,0)L=3,M\rangle&=&t^S_M,\,M={0,\pm 1}\nonumber\\
|(3,0)L=1,M\rangle&\equiv& T^{(30)1}_M=-\frac{2}{3}t^S_{M}+\sqrt{5}{3}t_M
\earr
\item  For the  $(\lambda,\mu)=(1,1)$  or octet $\underline{8}$ we have:
\barr
|(1,1)L=1,M\rangle&\equiv &T^{(11)1}_M=\frac{\sqrt{5}}{3}t^S_{M}+\frac{2}{3}t_M,\,M=\pm 1,0,\, |(1,1)L=2,M\rangle \equiv T^{(11)2}_M=t^A_{M},\,M=\pm 1,0,\nonumber\\
%|(1,1)L=2,M\rangle&\equiv &T^{(11)2}_M=t^A_{M},\,M=\pm 1,0,\, |(1,1)L=2,M\equiv T^{(11)2}_M=t^A_{M},\,M=\pm 1,0,\nonumber\\
|(1,1)L=2,M=2\rangle&=&\frac{1}{\sqrt{2}}\left(\underline{1}'+\underline{2}''\right),\,|(1,1)L=2,M=-2\rangle=\frac{1}{\sqrt{2}}\left(\underline{1}'-\underline{1}'' \right),
\earr
\item  For the  $(\lambda,\mu)=(0,0)$  or singlet $\underline{1}$ of $SU(3)$ we have:
\beq
|(00)L=0,M=0\rangle=\left [ \left [{\bf p}\otimes{\bf q}\right  ]^1\otimes{\bf r}\right ]^0_0=\underline{1}\mbox{ of }A_4
\eeq
\end{enumerate}
\section{The scalar potential}
We will restrict our discussion considering the  representations which contain only non invariant singlets, namely the octet $(\lambda,\mu)=(1,1) $  and the singlet $ (\lambda,\mu)=(0,0)$, which contain $A_4$ non invariant singlets. The second is written in the tensor decomposition as:
\beq
|(0,0)L=0, M=0\rangle=i [{\bf p}\times{\bf q}].{\bf r}
\eeq
It is an $A_4$ scalar which changes sign under conjuation and can be identified with $\underline{1}''$.
\\ 
The adjoined scalar representations is written as follows:
\beq
\phi^{(11)}_{LM}=\left ( \begin{array}{cc}L=1,M=1\Leftrightarrow&x_1=t^{((11),1)}_1\\ L=1,M=-1\Leftrightarrow&x_2=t^{((11),1)}_{-1}\\  L=1,M=0\Leftrightarrow&x_3=t^{((11),1)}_{0}\\ 
 L=2,M=2\Leftrightarrow&x_4=t^{((11),2)}_{1}\\ L=2,M=-2\Leftrightarrow&x_5=t^{((11),2)}_{-1}\\ L=2,M=0\Leftrightarrow&x_6=t^{((11),2)}_{0}\\
 L=2,M=2\Leftrightarrow&\frac{1}{\sqrt{2}}(x_7+x_8)\\ L=2,M=-2\Leftrightarrow &\frac{1}{\sqrt{2}}(x_7-x_8)\\
\end{array}
\right)
\label{Eq:phi11lm}
\eeq
with $x_7=\underline{1}'$ and $x_8=\underline{1}''$.

\barr
V&=&\sum_{i=1}^2(-m_1^2 X_i+\lambda_i X_i^2) +\lambda_{12}X_1 X_2+\sum_{i=3}^4\lambda_{i} [X_i\otimes X_i]^{(\lambda,\mu)=(0,0)L=0}+%\lambda_{34} [X_3\otimes X_4]^{(\lambda,\mu)=(0,0)L=0}
\nonumber\\
&+&\lambda_{34} [X_3\times X_4]^{(\lambda,\mu)=(0,0)L=0}+\sum_{i=5}^6 \lambda_{i}[X_i\otimes X_i]^{(\lambda,\mu)=(0,0)L=0}
+  \sum_{i=7,8,9}\lambda_i [X_i\otimes X_i]^{(\lambda,\mu)=(0,0)L=0}
\earr
%\beq
%V=\sum_{i=1}^2(-m_1^2 X_i+\frac{\lambda_i}{4} X_i^2) +\frac{\lambda_{12}}{4}X_1 X_2+\frac{\lambda_3}{4}V_{33}+ \sum_{i=5}^6 %\frac{\lambda_i}{4}V_{ii}+\frac{\lambda_{56}}{4}V_{56}
% +\sum_{i=7}^9\frac{\lambda_{i}}{4}V_{ii}
%\eeq
(in the notation of $X_i$ given in the appendix).
In other words the quartic terms of the potential are:
$$V_{ij} =[X_i\times X_j]^{(\lambda,\mu)=(0,0)L=0}$$ 
%Note that $V_{33}=V_{34}=0$\\
The term involving the singlet is trivial:
\beq
V_{11}=X_1^2=\left (\left (\underline{1}'\right )^2\right )^2
\eeq
 Focusing on the octet we have, e.g :
\beq 
X_2=-\frac{1}{\sqrt{8}}\left ( x_1 x_2+x_2 x_1 -x_3^2-x_4 x_5-x_5 x_4 +x_6^2+x_7^2-x_8 ^2\right ), V_{22}=X_2^2
\eeq
It is not necessary to explicitly construct the other terms $V_{i,j}$ since we are interested only in the terms involving the $A_4$ singlets. All of them lead to the same expression as above, namely $(X_7^2-x_8^2) 2$. In other words we have:
\barr
U&=&-m_1^2\left (x_7^2-x_8^2 \right)+\lambda_1 \left (x_7^2-x_8^2 \right)^2-\tilde{m}_2^2x_7^4+\lambda_2 x_7^4+\lambda_{12}x_8^2\left (x_7^2-x_8^2\right )\nonumber\\ &=&-m_1^2x_7^2+\lambda_1 \left (x_7^2-x_8^2 \right)^2-m_2^2x_7^2+\lambda_2 x_7^4+\lambda_{12}x_8^2\left (x_7^2-x_8^2\right ),\,m_2^2=\tilde{m}_2^2-m_1^2
\earr
This potential has a minimum at
\beq
\frac{\partial U}{\partial x_7}=0,\,\frac{\partial U}{\partial x_8}=0
\eeq
given by
\beq
x_7^2=v_a^2=\frac{\lambda _1 m_1^2+\lambda _2
   m_1^2-2 \lambda _3
   m_1^2+\lambda _1 m_2^2-\lambda
   _3 m_2^2}{\lambda _1 \lambda
   _2-\lambda _3^2},\,x_8^2=v_b^2=\frac{\lambda _1 m_1^2-\lambda _3
   m_1^2+\lambda _1
   m_2^2}{\lambda _1 \lambda
   _2-\lambda _3^2}
\eeq
provided that
$$
\left .\frac{\partial ^2 U}{\partial x_7^2}+\frac{\partial ^2 U}{\partial x_8^2}+2\frac{\partial ^2 U}{\partial x_7 x_8}\right |_{x_7=v_a,x_8=v_b}>0
$$
or
$$
\frac{4 \left(\lambda _1+\lambda
   _2-2 \lambda _3\right) \left(2
   \lambda _1-\lambda _3\right)
   m_1^2+4 \lambda _1 \left(2
   \lambda _1+\lambda _2-3
   \lambda _3\right)
   m_2^2}{\lambda _1 \lambda
   _2-\lambda _3^2}+$$
$$8 \left(\lambda _3-\lambda
   _1\right)
   \sqrt{\frac{\left(\lambda
   _1-\lambda _3\right)
   m_1^2+\lambda _1
   m_2^2}{\lambda _1 \lambda
   _2-\lambda _3^2}}
   \sqrt{\frac{\left(\lambda
   _1+\lambda _2-2 \lambda
   _3\right) m_1^2+\left(\lambda
   _1-\lambda _3\right)
   m_2^2}{\lambda _1 \lambda
   _2-\lambda _3^2}} >0
$$
This condition can be  satisfied for a wide range of the parameters.

After the singlets acquire a vacuum expectation value they get a mass, which is given after diagonalizing the matrix:
$$
\left(
\begin{array}{cc}
 \frac{4 \lambda _1
   \left(\left(\lambda _1+\lambda
   _2-2 \lambda _3\right)
   m_1^2+m_2^2 \left(\lambda
   _1-\lambda
   _3\right)\right)}{\lambda _1
   \lambda _2-\lambda _3^2} &
   \frac{4
   \sqrt{\left(\left(\lambda
   _1-\lambda _3\right)
   m_1^2+m_2^2 \lambda _1\right)
   \left(\left(\lambda _1+\lambda
   _2-2 \lambda _3\right)
   m_1^2+m_2^2 \left(\lambda
   _1-\lambda _3\right)\right)}
   \left(\lambda _3-\lambda
   _1\right)}{\lambda _1 \lambda
   _2-\lambda _3^2} \\
 \frac{4
   \sqrt{\left(\left(\lambda
   _1-\lambda _3\right)
   m_1^2+m_2^2 \lambda _1\right)
   \left(\left(\lambda _1+\lambda
   _2-2 \lambda _3\right)
   m_1^2+m_2^2 \left(\lambda
   _1-\lambda _3\right)\right)}
   \left(\lambda _3-\lambda
   _1\right)}{\lambda _1 \lambda
   _2-\lambda _3^2} & \frac{4
   \left(\left(\lambda _1-\lambda
   _3\right) m_1^2+m_2^2 \lambda
   _1\right) \left(\lambda
   _1+\lambda _2-2 \lambda
   _3\right)}{\lambda _1 \lambda
   _2-\lambda _3^2} \\
\end{array}
\right)
$$
	Once the symmetry is broken the gauge bosons acquire a mass.
	The generalized derivative is: 
	\beq
	D_{\mu}\phi_{\alpha}=\partial_{\mu}\phi_{\alpha}+i g A^{(i)}_{\mu} \sum_{\beta}\phi_{\beta}R(\alpha,\beta,i),\,R(\alpha,\beta,i)=\sum_{\rho}\langle(1,1)||T^{(1,1)}||(1,1)\rangle_{\rho} \langle (1,1)\ell_{\alpha}m_{\alpha};(1,1)\ell_i,m_i|(1,1)\ell_{\beta}m_{\beta}\rangle_{\rho}
	\eeq
	The reader should not be confused with the index $\mu$ which is a Lorentz index with the quantum numbers $L,M$, which, of course, have nothing to do with the angular momentum, but they refer to tensors in the flavor space.
	
	Thus\footnote{In the product $(11)\otimes (11)$ the representation (11) appears twice, i.e $\rho=1,2$ . The $\rho=1$ double bar coefficient is $\sqrt{3}$, while the other double bar coefficient can be set equal to zero \cite{HechtSU(3)}. So only the $\rho=1$  is needed}
	\barr
	\left(D^{\mu}\phi_{\alpha}\right)^+D_{\mu}\phi_{\alpha}&=&\partial ^{\mu}\left(\phi_{\alpha}\right)^+\partial _{\mu}\phi_{\alpha}+ig\left [\partial ^{\mu}\left(\phi_{\alpha}\right )^+A^{(i)}_{\mu} -\partial _{\mu}\phi_{\alpha}A^{(i)\mu} \right ]R(\alpha,\beta,i)\nonumber\\&+& g^2 A^{(i)}_{\mu}A^{(j)\mu} \sum_{\alpha,\beta,\beta'}\left(\phi_{\beta'}\right)^+\phi_{\beta}\,R(\alpha,\beta,i)R(\alpha,\beta',j)
	\earr
	Once the scalar fields  acquire a vacuum expectation value the gauge bosons acquire a mass. The mass matrix is given by:
	\beq
	{\cal M}^2=\frac{1}{2}m_{i,j}^2,\,m_{i,j}^2=g^2\sum_{\alpha,\beta,\beta'}R(\alpha,\beta,i) R(\alpha,\beta',j)\left . \left[\left(\phi_{\beta'}\right )^+\phi_{\beta}\right ]  \right |_{\left(\phi_{\beta'}\right )^+\phi_{\beta}=v'_{\beta}v_{\beta}}
	\label{Eq:gaugemass1}
	\eeq
	The coefficients can easily be evaluated.
	% The double bar coefficient is $\sqrt{3}$ it is of no consequence, since it is an overall normalization and it can be absorbed in the gauge coupling $g$. Hence  we find
\beq
	R(\alpha,\beta,i)= \sqrt{3}\langle (1,1)\ell_{\beta};(1,1)\ell_i|(1,1)\ell_{\alpha}\rangle \langle\ell_{\beta}m_{\beta};\ell_i,m_i|\ell_{\alpha}m_{\alpha}\rangle
	\eeq
	The first factor is the reduced double bar coefficient, the second  is the $SU(3)\supset S(O)$ C-G coefficient with the nonzero values given by the $X_3$ and $X_5$ tabulated in the appendix and  last  factor is just an ordinary C-G coefficient.  Since, however, the singlets are characterized by $L=2$ only the  $X_5$ expressions are relevant. Then the gauge boson mass matrix expressed in components $T^{(1,1)L}_M $, becomes:
	
	\beq
	\begin{array}{|c|cccc|c|}
	\hline
	&T^{(1,1)1}_1&T^{(1,1)1}_0&T^{(1,1)2}_2&T^{(1,1)2}_1\\
	\hline
\left (T^{(1,1)1}	_1 \right )^*&1&-\frac{1}{\sqrt{2}}&-\frac{1}{\sqrt{2}}&\frac{1}{2}\\
\left (T^{(1,1)1}	_0 \right )^*&-\frac{1}{\sqrt{2}}&2&1&0\\
\left (T^{(1,1)2}	_2  \right )^*&-\frac{1}{\sqrt{2}}&1&2&-\frac{1}{\sqrt{2}}\\
\left (T^{(1,1)2}	_1  \right )^*&\frac{1}{2}&0&-\frac{1}{\sqrt{2}}&1\\
	\hline
				\end{array}.
	\label{Eq:gaugemass2}
	\eeq
The above matrix is, of course, given in units of
$$ m^2_b= g^2\frac{\lambda _2 m_1^2-\lambda _3
   \left(m_1^2+m_2^2\right)}{\lambda _1 \lambda _2-\lambda
   _3^2}
$$
the above matrix has four non zero eigenvalues leading to 4 massive gauge bosons with  masses squared $(3.58,1.44,0.69,0.48)m^2_b$. The corresponding transformation is given by:
$$
\left (
\begin{array}{c}
T^{(1,1)1}_1\\T^{(1,1)1}_0\\T^{(1,1)2}_2\\T^{(1,1)2}_1\\T^{(1,1)2}_0\\ \end{array}
\right )
\left(
\begin{array}{ccccc}
 0.390671 & 0.105104 & 0.699252 &
   0.589386&0 \\
 -0.589386 & 0.699252 & -0.105104 &
   0.390671&0 \\
 -0.659264 & -0.317896 & 0.631618 &
   -0.255678&0 \\
 0.255678 & 0.631618 & 0.317896 &
   -0.659264 &0\\
   0&0&0&0&0
\end{array}
\right)\left (
\begin{array}{c}
W^{(0)}_1\\W^{(0)}_2\\W^{(0)}_3\\W^{(0)}_4\\W^{(0)}_5\\ \end{array}
\right )$$ 
The reason for the superscript (0) will become clear below.\\
So out of the  gauge bosons of $SU(3)$, three "charged" pairs and two "neutral", only one "neutral" remains massless. This  can be identified  with the $T^{(1,1)2}_0$, which did not appear in the above mass matrix,  Eq. (\ref{Eq:gaugemass2}). This means that  the corresponding generator remains unbroken, i.e. a symmetry has survived. Therefore some additional Higgs scalar must acquire a vacuum expectation value. We have chosen this to be the field $\Phi^{(1,1)1}_0$. Then proceeding as above we find one gauge boson acquires a mass of $2.1 m^2_{a}$ with $ m_a^2=g^2 \left( v^{(11)1}_0\right)^2$ with components $$\left (\sqrt{\frac{3}{14}},-\sqrt{\frac{2}{7}},0,-\sqrt{\frac{3}{14}},\sqrt{\frac{2}{7}}\right )$$ in the above basis. The five states that found constitute a basis in the gauge boson basis, but they are not orthogonal. They can be orthogonalized to lead to the states 
$$
\left (\begin{array}{c}
W_1\\W_2\\W_3\\W_4\\W_5\\
\end{array}\right )
\left(
\begin{array}{cccccc}
 -0.714625 & 0.389224 & -0.733933 &
   -0.0992339 & 0.0599715 \\
0.825178 & 0.741429 & -0.0492126 &
   -0.218771 & -0.0692491 \\
 0 & -0.284231 & -0.0211346 & -0.958523
   & 0 \\
 0.412793 & 0.412093 & -0.677107 &
   0.153381 & 0.24186 \\
-0.377964 & 0 & 0 & 0 & -0.377964 \\
\end{array}
\right)
\left (\begin{array}{c}
T^{(1,1)1}_1\\T^{(1,1)1}_0\\T^{(1,1)2}_2\\T^{(1,1)2}_1\\T^{(1,1)2}_0\\
\end{array}\right )
$$
with masses squared given as follows:
$$m_1^2=1.05 m_a^2+1.79 m_b^2,\,m_2^2=0.210 m_a^2+0.870 m_b^2,\,m_3^2=0.561 m_a^2+1.026 m_b^2,$$ $$m_4^2=0.080 m_a^2+1.114 m_b^2,\,m_5^2=0.649 m_a^2+2.943 m_b^2$$

\section{Discussion}
In this work we considered the flavor symmetry $SU_f(3)\supset SO(3)\supset A_4$. We have developed the formalism for constructing  in this chain all the $SU_f(3)$  invariants made up of the fundamental representation or octet ($\underline{8}$), the decuplet $\underline{10}$ and the singlet $\underline{1}$  $SU_f(3)$ representations. Then we examined the spontaneous  braking of this symmetry by allowing the non invariant $A_4$  singlet scalars contained in $\underline{8}$ and $\underline{1}$ to acquire a vacuum expectation value. After that we find that this is not adequate to make all the gauge bosons massive. This was achieved by allowing one component of one of the $A_4$ isotriplets contained in the $\underline{8}$ to also attain a vacuum expectation value. This resulted in all gauge bosons attaining a mass given as a function of the gauge coupling $g$ and the vacuum expectation values.

After this what remains to be done is to place all fermions of the theory into $SU_f(3)\supset SO(3)\supset A_4$ representations. Then using the above formalism it is easy to write down the invariant $SU_f(3)$ combinations of Yukawa couplings. Then after the breaking of the symmetry as mentioned above, one obtains fermion masses and mixings as well as gauge interactions involving the fermion mass eigenstates. 

{\bf Acknowledgments}: The author is indebted to Professors G. K. Leontaris for useful discussions and Tony Thomas for his hospitality and support during his visit in Adelaide.

\section{Appendix: The needed $SU(3)\supset SO(3)$  C-G coefficients}
The full C-G coefficient needed for the reduction of $SU(3)\otimes SU(3)$ representations factorizes into one $SU(3)\supset SO(3)$ and one $SO(2)\supset SO(2)$ factor. Symbolically:
$$\langle(\lambda_1,\mu_1), \kappa_1,L_1,M_1,\,\lambda_2,\mu_2,\kappa_2,L_2M_2|\lambda_3,\mu_3,\kappa_3,L_3M_3\rangle=\langle(\lambda_1,\mu_1), \kappa_1,L_1,\,\lambda_2,\mu_2,\kappa_2,L_2||\lambda_3,\mu_3,\kappa_3,L_3\rangle$$ $$ \langle L_1 M_1,L_2,M_2|L_3,M_3\rangle.$$
So, since the usual $SO(2)\supset SO(2)$ C-G coefficient is well known, we need consider only the double bar coefficients. 
 The only $SU(3)\supset SO(3)$  C-G coefficients, not found in the literature \cite{JDV68} are those involving the $(3,0)\otimes(0,3)$ reduction.
On this occasion we will consider the general reduction $$(\lambda,0)\otimes(0,3)\rightarrow (\lambda',\mu'),\,(\lambda',\mu')=(\lambda,3),\,(\lambda-1,2),\,(\lambda-2,1),\,(\lambda-3,0)$$
To this end we will use the build-up procedure \cite{JDV68}
$$\langle(\lambda,0)L,(0,3) \ell ||(\lambda',\mu')\kappa'L'\rangle U\left ((\lambda,0),(0,2),(\lambda',\mu')(0,1)1;(\lambda^{''},\mu{''}),(0,3)\right )=$$ $$\sum_{\kappa^{''}L^{''}\ell_1}U(L,\ell_1,L',1;L^{''}\ell)\langle(\lambda,0)L,(0,2) \ell_1 ||(\lambda^{''},\mu{''})\kappa^{''}L^{''}\rangle$$ $$\langle(0,2)\ell_1,(01)1||(03)\ell\rangle\langle(\lambda^{''},\mu{''})\kappa^{''}L^{''},(0,1)||(\lambda',\mu')\kappa'L'\rangle $$
where $U(L,\ell_1,L',1;L^{''}\ell) $ is the (unitary) Racah 6j function for the rotation group and $U\left ((\lambda,0),(0,2),(\lambda',\mu')(0,1);(\lambda^{''},\mu{''}),(0,3)\right )$ is an analogous function for the SU(3) group, which can be judiciously chosen to make the above expression as simple as possible. It can be computed by requiring the C-G coefficients of the left hand side to be suitably normalized. A convenient choice is 
$$(\lambda^{''},\mu{''})=(\lambda,2),\,(\lambda-1,1),\,(\lambda-2,0),\,(\lambda-2,0)\Leftrightarrow (\lambda^{'},\mu{'})=(\lambda,3),\,(\lambda-1,2),\,(\lambda-2,1),\,(\lambda-3,0).$$
Using Eq. (25a) of the above reference it can be cast in the form:
\barr
\langle(\lambda,0)L,(0,3) \ell ||(\lambda',\mu')\kappa'L'\rangle&& U\left ((\lambda,0),(0,2),(\lambda',\mu')(0,1);(\lambda^{''},\mu{''}),(0,3)\right )=(-1)^{L-L'+(2-\ell_1)/2}(-1)^{(\mu'-\lambda'+\lambda-3)/3}\nonumber\\ &&\sqrt{\frac{\mbox{dim}(\lambda',\mu')}{\mbox{dim}(\lambda,0)}}\sqrt{\frac{2 L+1}{2 L'+1}} \sum_{\kappa^{''}L^{''}\ell_1}\sqrt{\frac{5}{3}\frac{2 \ell_1+1}{2 \ell+1}}U(L,\ell_1,L',1;L^{''}\ell)\nonumber \\ &&\langle(\lambda^{''},\mu{''})\kappa^{''}L^{''},(2,0) \ell_1 ||(\lambda,0)L\rangle\nonumber\\ &&\langle(0,3)\ell,(1,0)1||(02)\ell_1\rangle\langle(\lambda',\mu')\kappa'L',(1,0)1||(\lambda^{''},\mu{''})\kappa^{''}L^{''}\rangle
\earr
with $\mbox{dim}(\lambda,\mu)=\frac{1}{2}(\lambda+1)(\mu+1)(\lambda+\mu+2)$  the dimension of $(\lambda,\mu)$. All the $SU(3)$ G-C entering the right hand side of the previous equation can be found in the tables \cite{JDV68}. The parameter $\kappa$ needed in  case there exists degeneracy in $L$, will be omitted in the cases that such a degeneracy is not present. We have obtained analytic expressions for the results, but their quite cumbersome. So we will present here the results needed here. We should mention that only the entries with non zero valus of the C-G are included in the tables below.
$$(3,0)L,(03)\ell||(0,0)L'=0\rangle=\left \{\begin{array}{cc}\sqrt{\frac{3}{10}}&L=\ell=1\\-\sqrt{\frac{7}{10}}&L=\ell=3\\ \end{array} \right .\,;
(3,0)L,(03)\ell||(1,1)L'=1\rangle=\left \{\begin{array}{ccc}L&\ell&C-G\\ 1&1&\sqrt{\frac{14}{15}}\\3&3&-\sqrt{\frac{1}{15}}\\ \end{array} \right .
$$
$$
(3,0)L,(03)\ell||(1,1)L'=2\rangle=\left \{\begin{array}{ccc}L&\ell&C-G\\ 1&1&-\frac{3}{5}\sqrt{\frac{7}{5}}\\1&3&\frac{2}{5}\sqrt{\frac{7}{5}}\\ 3&1&\frac{2}{5}\sqrt{\frac{7}{5}}\\3&3&-\frac{1}{5}\sqrt{\frac{6}{5}}\\ \end{array} \right .\,;3,0)L,(03)\ell||(2,2)\kappa'=0,L'=4\rangle=\left \{\begin{array}{ccc}L&\ell&C-G\\ 3&3&-\sqrt{\frac{11}{35}}\\1&3&2\sqrt{\frac{3}{35}}\\ 3&1&-2\sqrt{\frac{3}{35}}\\ \end{array} \right .
$$
$$
(3,0)L,(03)\ell||(2,2)\kappa'=0,L'=2\rangle=\left \{\begin{array}{ccc}L&\ell&C-G\\ 3&3&\frac{12}{5}\sqrt{\frac{2}{91}}\\1&3&-\frac{1}{5}\sqrt{\frac{39}{7}}\\ 3&1&\frac{22}{5}\sqrt{\frac{3}{91}}\\1&1&\frac{2}{5}\sqrt{\frac{1}{3}}\\ \end{array} \right .\,;(3,0)L,(03)\ell||(2,2)\kappa'=2,L'=2\rangle=\left \{\begin{array}{ccc}L&\ell&C-G\\ 3&3&-\frac{12}{5}\sqrt{\frac{6}{65}}\\1&3&-\frac{2}{5}\sqrt{\frac{13}{5}}\\ 3&1&-\frac{2}{5}\sqrt{\frac{21}{65}}\\1&1&\frac{1}{5}\sqrt{\frac{1}{65}}\\  \end{array} \right .
$$
$$
(3,0)L,(03)\ell||(2,2)\kappa'=2,L'=3\rangle=\left \{\begin{array}{ccc}L&\ell&C-G\\ 3&3&{3}\sqrt{\frac{3}{35}}\\1&3&-{2}\sqrt{\frac{1}{35}}\\ 3&1&{2}\sqrt{\frac{1}{35}}\\ \end{array} \right .\,;(3,0)L,(03)\ell||(2,2)\kappa'=0,L'=0\rangle=\left \{\begin{array}{ccc}L&\ell&C-G\\ 3&3&-\sqrt{\frac{3}{10}}\\1&1&-\sqrt{\frac{7}{10}}\\  \end{array} \right .
$$
$$
(3,0)L,(03)\ell||(3,3)\kappa'=1,L'=1\rangle=\left \{\begin{array}{ccc}L&\ell&C-G\\ 1&1&{-}\sqrt{\frac{14}{15}}\\3&3&-\sqrt{\frac{1}{15}}\\ \end{array} \right .\,;(3,0)L,(03)\ell||(3,3)\kappa'=1,L'=2\rangle=\left \{\begin{array}{ccc}L&\ell&C-G\\ 1&1&\frac{2 \sqrt{2}}{5}\\ 1&3&-\frac{2}{5} \sqrt{\frac{6}{5}}\\ 3&1&\frac{2}{5} \sqrt{\frac{6}{5}}\\ 3&3&\frac{1}{5} \sqrt{\frac{1}{7}}\\   \end{array} \right .
$$
$$
(3,0)L,(03)\ell||(3,3)\kappa'=1,L'=3\rangle=\left \{\begin{array}{ccc}L&\ell&C-G\\ 1&3&{-}\sqrt{\frac{3}{55}}\\3&1&-4\sqrt{\frac{3}{55}}\\3&3&2\sqrt{\frac{1}{55}}\\ \end{array} \right .\,;(3,0)L,(03)\ell||(3,3)\kappa'=3,L'=3\rangle=\left \{\begin{array}{ccc}L&\ell&C-G\\ 1&3&{-8}\sqrt{\frac{1}{77}}\\3&1&1\sqrt{\frac{1}{77}}\\3&3&-2\sqrt{\frac{3}{77}}\\ \end{array} \right .
$$
$$
(3,0)L,(03)\ell||(3,3)\kappa'=1,L'=4\rangle=\left \{\begin{array}{ccc}L&\ell&C-G\\ 1&3&{3}\sqrt{\frac{11}{287}}\\3&1&4\sqrt{\frac{11}{287}}\\3&3&-2\sqrt{\frac{3}{287}}\\ \end{array} \right .\,;(3,0)L,(03)\ell||(3,3)\kappa'=3,L'=4\rangle=\left \{\begin{array}{ccc}L&\ell&C-G\\ 1&3&{8}\sqrt{\frac{1}{205}}\\3&1&-3\sqrt{\frac{1}{205}}\\3&3&2\sqrt{\frac{33}{205}}\\ \end{array} \right .
$$
$$
(3,0)L,(03)\ell||(3,3)\kappa'=1,L'=5\rangle=-1\,;(3,0)L,(03)\ell||(3,3)\kappa'=1,L'=6\rangle=1
$$
Incidentally the $SU(3)$ U-functions are:
$$U((30)(02)(\lambda',\mu')(01);(\lambda^{''},\mu^{''})(03)=1,\frac{1}{\sqrt{2}},\frac{\sqrt{7}}{3},1$$ 
for
$$ (\lambda',\mu'),(\lambda^{''},\mu^{''})=(00),(1,0);\,(1,1),(1,0);\,(2,2)(2,1);\,(3,3)(3,2) $$
respectively.

We will now consider the self adjoined representations\\
In the case of the SU(3) scalar representation we have the quadratic invariance :
$$X_1=|(0,0)|^2$$
In the case of the $\underline{8}=(11)$ representation the needed $SU(3)$ C-G   can be obtained from the existing  tables \cite{JDV68}, but, for the reader's convenience, we will include them here.
$$X_2\Leftrightarrow(1,1)L,(11)\ell||(0,0)L'=0\rangle=\left \{\begin{array}{cc}-\sqrt{\frac{3}{8}}&L=\ell=1\\ \sqrt{\frac{5}{8}}&L=\ell=2\\ \end{array} \right .$$
$$
X_3\Leftrightarrow(1,1)L,(11)\ell||(1,1),L'=1\rangle_1=\left \{\begin{array}{ccc}L&\ell&C-G\\ 1&1&-\sqrt{\frac{1}{6}}\\2&2&\sqrt{\frac{5}{6}}\\ \end{array} \right .\,;X_4\Leftrightarrow(1,1)L,(1,1)\ell||(1,1)L'=1\rangle_2=\left \{\begin{array}{ccc}L&\ell&C-G\\ 2&1&\sqrt{\frac{1}{2}}\\1&2&\sqrt{\frac{1}{2}}\\ \end{array} \right .
$$
 $$
X_5\Leftrightarrow(1,1)L,(11)\ell||(1,1),L'=2\rangle_1=\left \{\begin{array}{ccc}L&\ell&C-G\\ 2&1&\sqrt{\frac{1}{2}}\\1&2&-\sqrt{\frac{1}{2}}\\ \end{array} \right .\,;X_6\Leftrightarrow(1,1)L,(1,1)\ell||(1,1)L'=2\rangle_2=\left \{\begin{array}{ccc}L&\ell&C-G\\ 1&1&-\sqrt{\frac{3}{10}}\\2&2&-\sqrt{\frac{7}{10}}\\ \end{array} \right .
$$
 $$
(1,1)L,(11)\ell||(3,0),L'=3\rangle=\left \{\begin{array}{ccc}L&\ell&C-G\\ 2&1&-\sqrt{\frac{1}{4}}\\1&2&\sqrt{\frac{1}{4}}\\2&2&\sqrt{\frac{1}{2}}\\ \end{array} \right .\,;(1,1)L,(1,1)\ell||(3,0)L'=1\rangle=\left \{\begin{array}{ccc}L&\ell&C-G\\ 2&1&\sqrt{\frac{1}{2}}\\1&2&\sqrt{\frac{1}{2}}\\ \end{array} \right .
$$
 $$
(1,1)L,(11)\ell||(0,3),L'=3\rangle=\left \{\begin{array}{ccc}L&\ell&C-G\\ 2&1&-\sqrt{\frac{1}{4}}\\1&2&\sqrt{\frac{1}{4}}\\2&2&-\sqrt{\frac{1}{2}}\\ \end{array} \right .\,;(1,1)L,(1,1)\ell||(0,3)L'=1\rangle=\left \{\begin{array}{ccc}L&\ell&C-G\\ 2&1&\sqrt{\frac{1}{2}}\\1&2&-\sqrt{\frac{1}{2}}\\ \end{array} \right .
$$
 $$
X_7\Leftrightarrow(1,1)L,(11)\ell||(2,2),L'=0\rangle=\left \{\begin{array}{ccc}L&\ell&C-G\\ 1&1&-\sqrt{\frac{5}{8}}\\2&2&\sqrt{\frac{3}{8}}\\ \end{array} \right .;\,X_8\Leftrightarrow (1,1)L,(1,1)\ell||(2,2)L'=3\rangle=\left \{\begin{array}{ccc}L&\ell&C-G\\ 2&1&\sqrt{\frac{1}{2}}\\1&2&\sqrt{\frac{1}{2}}\\ \end{array} \right .
$$
 $$
(1,1)L,(11)\ell||(2,2),\kappa'=0,L'=2\rangle=\left \{\begin{array}{ccc}L&\ell&C-G\\ 1&1&-\frac{1}{2}\sqrt{\frac{7}{13}}\\2&1&-\frac{1}{2}\sqrt{\frac{21}{13}}\\  1&2&\frac{1}{2}\sqrt{\frac{21}{13}}\\2&2&\frac{1}{2}\sqrt{\frac{3}{13}}\\ \end{array} \right .\,;(1,1)L,(1,1)\ell||(2,2)\kappa'=2,L'=2\rangle=\left \{\begin{array}{ccc}L&\ell&C-G\\ 1&1&\frac{7}{2}\sqrt{\frac{3}{65}}\\2&1&-\frac{1}{2}\sqrt{\frac{5}{13}}\\  1&2&\frac{1}{2}\sqrt{\frac{5}{13}}\\2&2&-\frac{3}{2}\sqrt{\frac{7}{65}}\\ \end{array} \right .
$$
 $$
X_9\Leftrightarrow(1,1)L,(11)\ell||(2,2),L'=4\rangle=1$$

\end{document}